\documentclass[aps,prl,reprint,superscriptaddress,amsmath,amssymb]{revtex4-2}

\usepackage{graphicx}
\usepackage{dcolumn}
\usepackage{bm}
\usepackage{bbm}
\DeclareMathOperator{\Tr}{tr}

\begin{document}


\title{The Effect of Confinement on Capillary Phase Transition In Granular Aggregates}


\author{Siavash Monfared}
\email{monfared@caltech.edu}
\affiliation{Division of Engineering and Applied Science, California Institute of Technology, Pasadena, CA 91125, USA.}
\altaffiliation[Previously at ]{Department of Civil and Environmental Engineering, Massachusetts Institute of Technology, Cambridge, MA 02139, USA.}

\author{Tingtao Zhou}
\email{tingtaoz@caltech.edu}
\affiliation{Division of Engineering and Applied Science, California Institute of Technology, Pasadena, CA 91125, USA.}
\altaffiliation[Previously at ]{Department of Physics, Massachusetts Institute of Technology, Cambridge, MA 02139, USA.}

\author{Jos\'{e} E. Andrade}
\email{jandrade@caltech.edu}
\affiliation{Division of Engineering and Applied Science, California Institute of Technology, Pasadena, CA 91125, USA.}

\author{Katerina Ioannidou}
\email{aikaterini.ioannidou@umontpellier.fr}
\affiliation{CNRS, University of Montpellier, LMGC, 163 rue Auguste Broussonnet
F-34090 Montpellier, France.}
\affiliation{MultiScale Material Science for Energy and Environment UMI 3466 CNRS-MIT-Aix-Marseille Universi\'{t}e Joint Laboratory, Cambridge, MA 02139, USA.}

\author{Farhang Radja\"{i}}
\email{franck.radjai@umontpellier.fr}
\affiliation{CNRS, University of Montpellier, LMGC, 163 rue Auguste Broussonnet
F-34090 Montpellier, France.}

\author{Franz-Josef Ulm}
\email{ulm@mit.edu}
\affiliation{Department of Civil and Environmental Engineering, Massachusetts Institute of Technology, Cambridge, MA 02139, USA.}

\author{Roland J.-M. Pellenq}
\email{Roland.pellenq@cnrs.fr}
\affiliation{Department of Physics, Georgetown University, Washington, D.C. 20057, USA.}
\affiliation{MultiScale Material Science for Energy and Environment UMI 3466 CNRS-MIT-Aix-Marseille Universi\'{t}e Joint Laboratory, Cambridge, MA 02139, USA.}

\date{\today}

\begin{abstract}
Utilizing a 3D mean-field lattice-gas model, we analyze the effect of confinement 
on the nature of capillary phase transition in granular aggregates with varying disorder 
and their inverse porous structures obtained by interchanging particles and pores. 
Surprisingly, the confinement effects are found to be much less pronounced 
in granular aggregates as opposed to porous structures. We show that 
this discrepancy can be understood in terms of the surface-surface correlation length with 
a connected path through the fluid domain, suggesting that this length captures 
the true degree of confinement. We also find that the liquid-gas phase transition 
in these porous materials is of second order nature near capillary critical 
temperature, which is shown to represent a true critical temperature, 
{\it i.e.} independent of the degree of disorder and the nature 
of solid matrix, discrete or continuous. 
The critical exponents estimated here from finite-size scaling analysis suggest 
that this transition belongs to the 3D random field Ising model universality class as 
hypothesized by P.G. de Gennes, with the underlying random fields induced by 
local disorder in fluid-solid interactions.
\end{abstract}

\maketitle

The fluid behavior confined in a solid matrix is of interest to a range of scientific and 
engineering fields, including wet granular physics and poromechanics \cite{Bocquet1998,Scheel2008,Coussy2010}, 
plant biology \cite{Ghestem2011,Radjai2019}, carbon capture 
technologies \cite{Snbjrnsdttir2020}, catalysis \cite{Davis2002,Czaja2009} and 
optics \cite{Barthelemy2007}. The behavior of a confined fluid contrasts significantly with that of a bulk fluid. This is a consequence of pore morphology, topology 
and the relative strength of fluid-solid to fluid-fluid interactions that alter the 
energy landscape of a 
fluid \cite{deGennes1985a,Thommes1994,Gelb1999,Pellenq2002,Pellenq2003,Pellenq2008,Barsotti2018,Parry2020}. In particular, the degree to which a fluid experiences confinement results in a shifted liquid-gas phase transition \cite{Evans1990,Binder1992,Gelb1999}. 
This effect is best captured through the concept of {\it capillary criticality} that 
hinges on the existence of a temperature $T_{cc}$ below the bulk critical temperature  
beyond which liquid-gas phase transition becomes reversible. 

For disordered porous materials, the nature of liquid-gas phase transition and 
whether capillary criticality is associated with a true critical point, i.e. termination of the liquidus line, 
are still unclear \cite{Monson2006,Evans2007}. Additionally, a central issue is how the effective random fields induced by structural and/or chemical disorder affect the 
degree of confinement, critical exponents and thus universality class classification. 
Bulk liquid-gas phase transitions are generally in the same universality class as the Ising 
ferromagnet \cite{Yang1952,Stanley1971}. It was conjectured by P.G. de Gennes that the universality class 
of liquid-gas phase transition in disordered porous materials should be that of the random-field Ising model (RFIM) \cite{deGennes1983,deGennes1984}. His argument is built on the stochastic nature of effective 
wall separation in disordered porous media that manifests itself as a quenched random variable in space. 

Inspired by analogies between jammed granular packings and disordered porous solids 
highlighted recently via studies on mechanics of dry systems \cite{Behringer2011,Laubie2017C}, 
we explore in this Letter the capillary phase transition in granular aggregates (discrete) and their inverse 
porous structures obtained by interchanging pores and particles (continuous). Based on extensive lattice-gas 
simulations, we examine 1) whether $T_{cc}$ represents a true critical temperature, 
2) the nature of phase transition as $T\rightarrow T_{cc}$, and 3) de Gennes'  
hypothesis \cite{deGennes1984} that critical behavior of fluids in random porous media 
can be mapped into the RFIM \cite{Ma1975} universality class. This has been only confirmed 
in colloid-polymer mixtures confined in random porous media and via Monte Carlo 
simulations \cite{Vink2006,Vink2008,Vink2009}. As we shall see, the confinement effects differ in the 
two types of structure, but in both cases de Gennes' hypothesis holds and  $T_{cc}$ appears to 
be a true critical temperature. 

Let us consider a set of granular media (GM) composed of rigid, non-overlapping monodisperse 
spherical particles, each confined to a cubic box of size $L_{x}=L_{y}=L_{z}=80$ nm with 
a reservoir of length $L_{\text{res.}}=5$ nm added in all directions. 
The first three structures $A$, $B$, $C$, each consists of $N_{p}=512$ particles with radius $R=4.7$ nm, have a packing fraction $f_{s}=0.43$, 
but exhibiting contrasting pore size, $r_{p}$, distributions (PSD) and 
increasingly more spatial disorder. 
Structure $D$, $N_{p}=955$ and $R=4$ nm, exhibits similar degree of 
spatial disorder as structure $C$ but with $f_{s}=0.5$ (see Supplemental Material for porous structure generation and PSD characterization \cite{SIPRL2020}). The corresponding inverse or \textit{negative} structures are porous solids (PS) 
obtained by switching pores and particles. We also consider a set of cylindrical pores (CP) of length $L_{x}=160$ nm $\ll L_{y}=L_{z}$ with pore radius $r_{p}\in\{2, 4, 8\}$
nm and with reservoirs of length $L_{\text{res.}}=4$ nm added to both ends.

\begin{figure}[t!]
\centering
\includegraphics[trim={0.0cm 0.0cm 0.0cm 0.0cm},clip,width=2.6in]{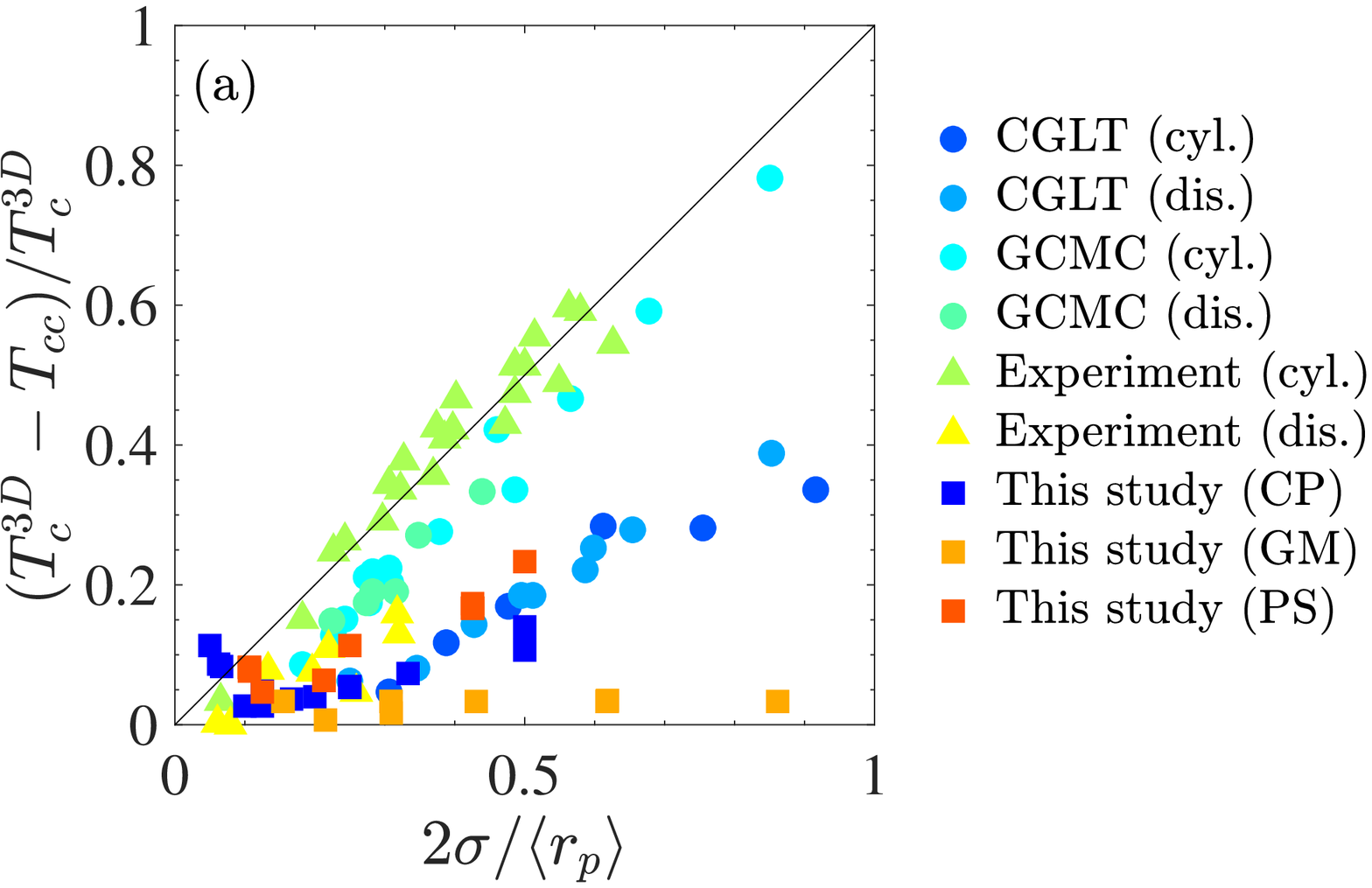}
\includegraphics[trim={0.2cm 0.25cm 0.3cm 0.2cm},clip,width=3.4in]{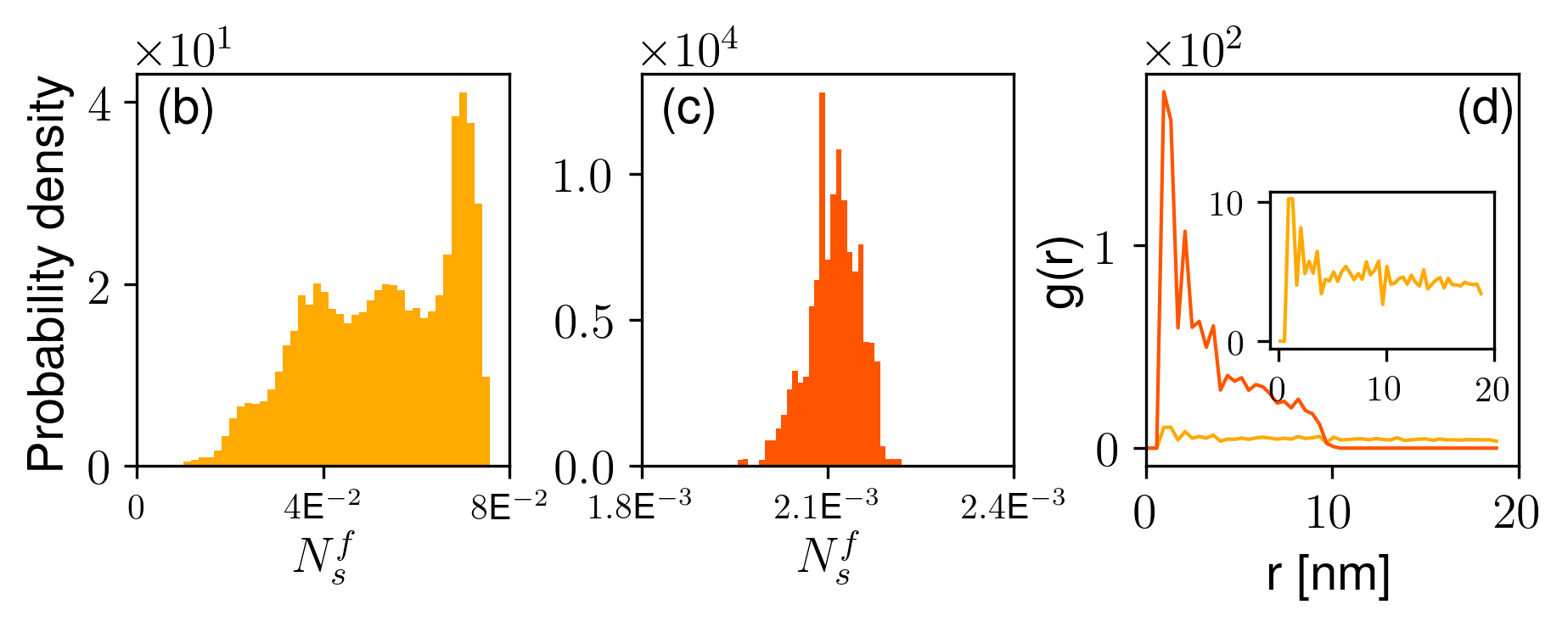}
\caption{(a) Shift in capillary critical temperature $T_{cc}$ as a function 
of the ratio $2\sigma/\langle r_{p}\rangle$. 
Previous simulations (circle) and experimental (triangle) data from literature  \cite{Pellenq2013}
along with our results (square) for all the considered cylindrical pores (CP), porous 
solids (PS) and granular media (GM) and for various space discretizations $a_{0}$, 
where $a_{0}\sim\sigma$. (b) and (c)  $N_{s}^{f}\left(r=20\text{ nm}\right)$ of GM and PS, respectively, 
for structure $C$. (d) Partial radial distribution function for a fluid site at the pore-solid interface 
with a connected path to a solid site at the pore-solid interface.}
\label{fig:2}
\end{figure}

\begin{figure*}[t!]
\centering
\includegraphics[trim={0.2cm 0.3cm 0.3cm 1.0cm},clip,width=4.8in]{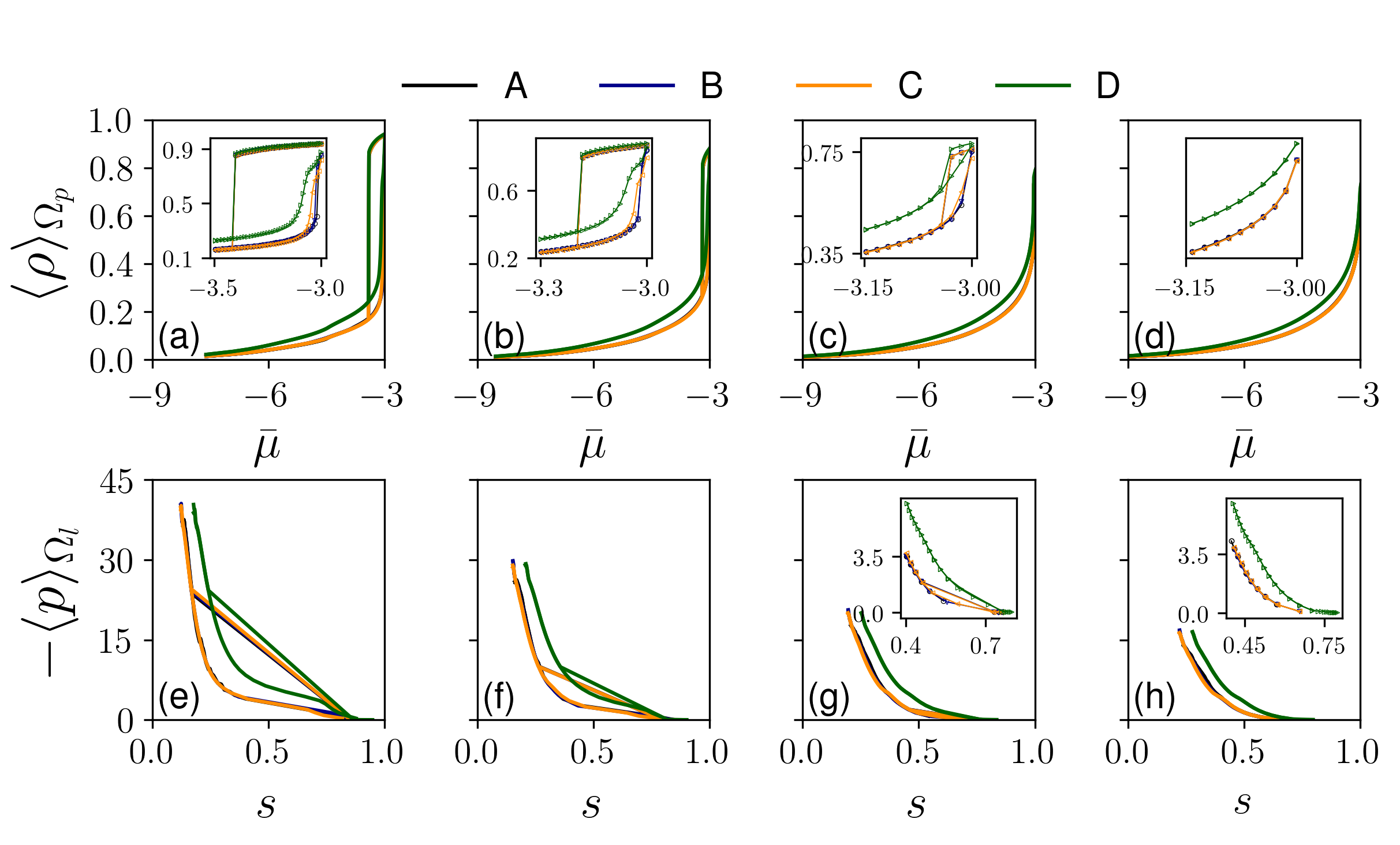}
\caption{Isotherms (a)-(d) and capillary pressure curves (e)-(h) for granular packings at $\bar{T}=1.0$, $\bar{T}=1.2$, $\bar{T}=1.4$ and $\bar{T}=1.5$, respectively.}\label{fig:4}
\end{figure*}

We use a parallelized implementation of Coarse-Grained Lattice 
gas density functional Theory (CGLT) \cite{Kierlik2001,Kierlik2002} with 
periodic boundary conditions in all directions on a simple cubic lattice with coordination 
number $c=6$. In this mean-field approach, the fluid is modeled in the grand canonical
ensemble via minimizing the grand potential $\Omega$ with normalized density field 
$\rho\left(\vec{x}\right)$ serving as the only order parameter in the model:
\begin{eqnarray}
\label{eq1}
\Omega &=& -w_{ff}\sum_{\langle i,j\rangle}\rho_{i}\rho_{j}-w_{sf}\sum_{i,j}\rho_{i}\eta_{j}-\mu\sum_{i}\rho_{i}+\notag \\ 
&& k_{B}T\sum_{i}[\rho_{i}\ln\left(\rho_{i}\right)+\left(1-\rho_{i}\right)\ln\left(1-\rho_{i}\right)], 
\end{eqnarray}
where $\eta_{i}=0 ( =1)$ indicates occupancy of site $i$ with solid (fluid). 
$w_{ff}$ and $w_{sf}$ represent fluid-fluid and fluid-solid energy interaction 
parameters where $y=w_{sf}/w_{ff}$ is set to $y=2.5$, 
corresponding to a strong fluid-solid surface affinity akin to methane in 
porous carbon or water in cement \cite{Bonnaud2012}. 

Based on our lattice choice, the normalized bulk critical temperature 
$\bar{T}_{c}^{3D}=k_{B}T^{3D}_{c}/w_{ff}=c/4=1.5$, 
and the normalized chemical potential corresponding to bulk liquid-gas phase transition 
$\bar{\mu}_{\text{sat}}^{3D}=\mu_{\text{sat}}^{3D}/w_{ff}=-c/2=-3$ are set. 
In the continuum limit and with correct parameterization, CGLT (Eq.\eqref{eq1}) 
approaches the Cahn-Hilliard model \cite{Hilliard1958,Zhou2019a}, 
paving the way to capture the liquid-gas interface diffusively \cite{Gibbs1876,Rayleigh1892}. 
This provides access to capillary stresses as a tensorial field, $\bm\sigma\left(\vec{x}\right)$, 
via Kortweg stress definition \cite{Kortweg1901,Anderson1997} and 
subsequently capillary pressure scalar field, $p\left(\vec{x}\right)=\left(1/3\right)\Tr\bm\sigma\left(\vec{x}\right)$ (see Supplemental Material \cite{SIPRL2020}):
\begin{equation}
\label{eq2}
\bm\sigma=\left(p_{0}\left(\rho\right)-\frac{\kappa}{2}\left(\vec{\nabla}\rho\right)^{2}\right)\bm{I}+\kappa\vec{\nabla}\rho\otimes\vec{\nabla}\rho+\bm{\sigma}_{0},
\end{equation}
where $p_{0}\left(\rho\right)=\mu\rho+\left(c w_{ff}/2\right)\rho^{2}-k_{B}T[\rho\ln\left(\rho\right)+\left(1-\rho\right)\ln\left(1-\rho\right)]
$ is the asymptotic bulk value of the hydrostatic pressure, $\bm{I}$ is the identity tensor, 
$\bm\sigma_{0}$ represents an arbitrary constant tensor, $\kappa=a_{0}^{2}w_{ff}$, 
and $a_{0}$ denotes lattice spacing. For proper energy scaling in this mean-field approach, 
$a_{0}$ is determined from liquid-gas surface tension. 

\begin{figure}[t!]
\centering
\includegraphics[trim={0.1cm 0.7cm 0.1cm 0.1cm},clip,width=3.4in]{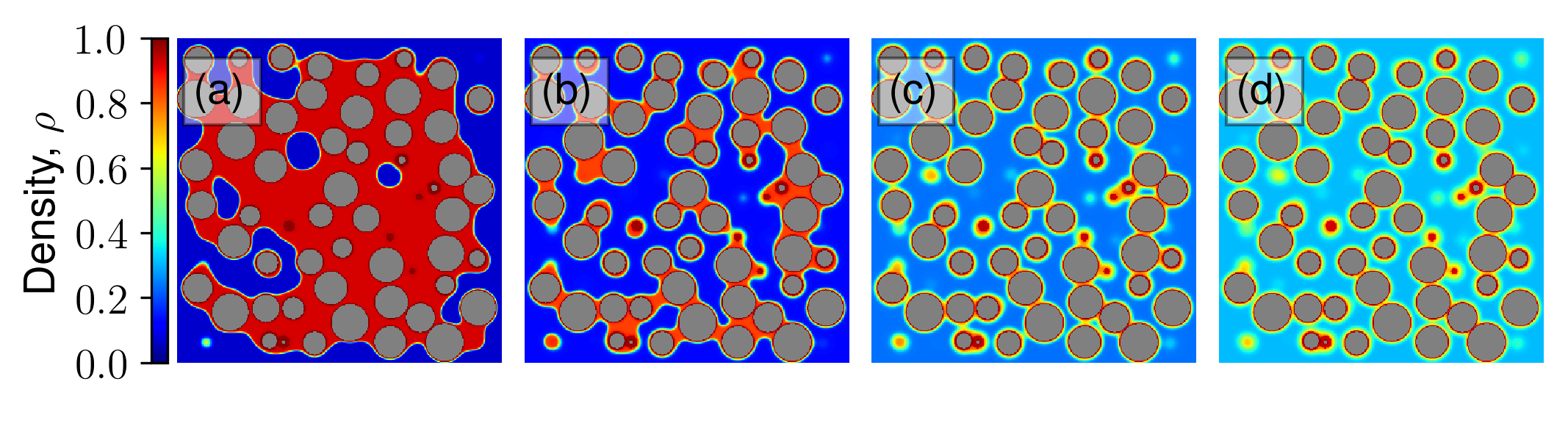}
\includegraphics[trim={0.1cm 0.12cm 0.2cm 0.25cm},clip,width=3.4in]{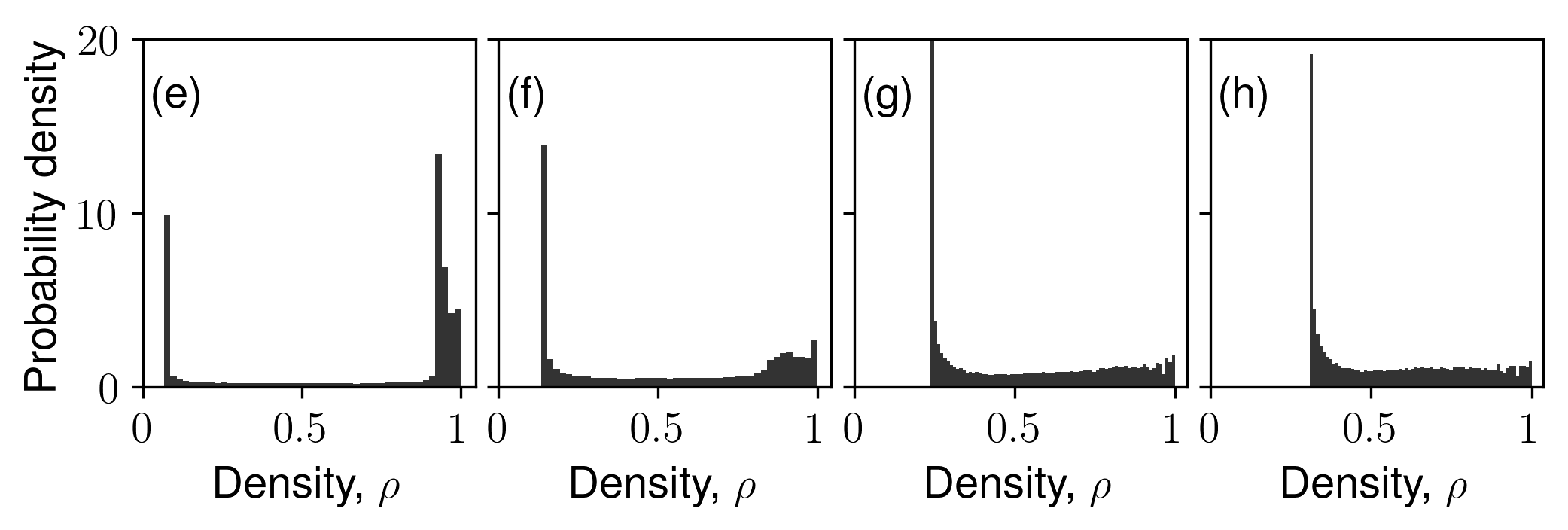}
\caption{(a)-(d): Spatial distribution of density fields, $\rho\left(\vec{x}\right)$, for granular packing $C$ and at cross-sections corresponding to $z=44$ nm, $h=0.96$ and (e)-(h): Probability density for the density fields (coarse-grained over $L^{3}=1$ nm$^{3}$) for the same structure at $h=0.96$ for $\bar{T}=1.0$, $\bar{T}=1.2$, $\bar{T}=1.4$ and $\bar{T}=1.5$, respectively.}\label{fig3}
\end{figure}

Figure 1(a) displays the calculated shift in critical temperature for GM, PS and CP as a function of the ratio $2\sigma/\langle r_{p}\rangle$ 
with $\sigma$ denoting the characteristic diameter of a fluid molecule, 
set equal to lattice spacing $a_0$. It also shows the data from experiments   
\cite{Kruk1999,Thommes1994,Burgess1990} and previous simulations based on CGLT and Grand 
Canonical Monte Carlo (GCMC) for cylindrical pores \cite{Salazar2005,Brovchenko2004,Pellenq2007}, 
e.g. MCM-41, carbon nanotube, disordered porous solids, i.e. Vycor and for a variety of 
substances, e.g. xenon, argon and water.   
The one-to-one scaling between the shift in capillary critical temperature and inverse 
mean pore size suggests similar behaviors independently of fluid and solid properties 
and pore connectivity. In this vein, we determine $T_{cc}$ with a resolution 
of $\approx 4$ K for CP, GM and PS with a lattice spacing 
$a_{0}\in\{0.2,0.25,0.5,1\}$ nm for CP and $a_{0}\in\{0.25,0.5,1\}$ nm for 
both GM and PS. For CP and PS, our results are in full agreement with 
the data reported in the literature \cite{Pellenq2013} and shown in Fig. 1(a). 

The CGLT is a mean-field approach that ignores thermal fluctuations and here is based on nearest neighbor interactions. This differs from the GCMC approach that accounts 
for thermal fluctuations leading to a lower $T_{cc}$ by $\approx 10$ K. 
This is what we observe here for CP compared to those in the literature based on 
GCMC and CGLT. However, the confinement imposed by GM seems insignificant and 
independent of the degree of spatial disorder, since $T_{cc}\lessapprox T^{3D}_{c}$. 
Let us explore this contrast further. 

The PSD in each considered PS has a peak 
that corresponds to the monodisperse particle radii with no variations around this peak. 
For the porous solids reported in the literature such as Vycor, the PSD can be captured 
by a Gaussian fit with a well-pronounced peak representing the mean and a small 
variance around it \cite{Levitz1991}. However, the PSD for GM exhibits a wide range 
and is not well represented by the first moment of the distributions (see Supplemental Material \cite{SIPRL2020}). 
To further characterize these distributions, we consider the proportion $N_{s}^{f}\left(r\right)$ 
of interface solid sites in a spherical 
domain of radius $r$, assuming that each site affects the evolution of a given interface fluid site through a connected path in the fluid domain, and 
normalized by the total number of interface solid sites. 
$N_{s}^{f}\left(r\right)$ represents the range of fluid-fluid correlations that 
can develop from the pore surface. Therefore, it 
contains information regarding correlation length for the adsorbed fluid or 
surface-surface correlation length. The distributions of $N_{s}^{f}\left(r=20\text{ nm}\right)$ 
as shown in Figs. 1(b)-1(c) for structure $C$ highlight the difference in 
confinement experienced by a fluid site in a granular material as opposed to a porous solid. 
For PS, each distribution is a Gaussian with a sharp peak at the mean and a 
small variance around it while for GM, they are no longer Gaussian, but  
distributed widely and multi-modal with the largest peak having a lower probability 
density than their porous solid counterparts. 
For PS, these distributions imply that every fluid site at the pore-solid interface 
has a high probability of interacting with a fixed number of solid sites while 
this probability is lower and the number of such interactions more widespread for GM. This notion is also reiterated in the partial radial distribution 
functions for fluid sites at the pore-solid interface interacting with solid sites as shown in Fig. 1(d). 

Thus, this disparity in adsorbed fluid correlation or surface-surface 
correlation length emerging from \textit{switching} solid curvature leads to more 
pronounced confinement effects in PS as opposed to their GM counterparts 
for which surface-surface correlation length approaches fluid-fluid correlation 
length in the bulk. Our results seem to depart from the Monte Carlo based 
study reported in \cite{Monson1996} for disordered granular packings with a similar packing 
fraction $f_{s}=0.386$ as our study for which a more pronounced shift in critical 
temperature is observed. This can be attributed to space discretization since in our study, 
the size ratio between the solid particle diameter ($2R$) and a fluid molecule ($\sigma$) 
is $\approx 40:1$ while in \cite{Monson1996} this ratio is $7.055:1$. 
Thus, our results suggest that bulk fluid behavior prevails in granular media 
exhibiting at least $2R/\sigma\gtrapprox 40$.

We now turn our focus to capillary pressure fields inside the pore domain, 
$\Omega_{p}$, defined as all fluid sites with no solid neighbors. 
Prior to any analyses, the average pressure of the reservoir is subtracted from 
the capillary pressure field $p\left(\vec{x}\right)$. 
The lattice spacing is chosen based on water-air surface tension $\gamma_{\text{lg}}\approx w_{ff}/2a_{0}\approx 72$ mN/m 
at $T=300$ K and thus $a_{0}\sim 0.25$ nm \cite{Zhou2019a}, 
comparable to the size of a water molecule. Having determined $T_{cc}$ associated with 
these physical parameters and for the granular aggregates considered previously, 
we simulate capillary condensation and evaporation for $\bar{T}=k_{B}T/w_{ff}\in\{1.0,1.2,1.4,1.5\}$ 
with the corresponding adsorption and desorption isotherms, as shown in Figs. 2(a)-2(d). 
The hysteresis loop is present for $\bar{T}\leq 1.4$ but it disappears at $\bar{T}=1.5$ 
with its shape becoming less symmetric with increasing temperature, 
a signature of disordered porous materials. Similar observation regarding the 
disappearance of the hysteresis loop can be made for CP (see Supplemental Material \cite{SIPRL2020}). Furthermore, there is a jump in mean density at $\bar{T}=1.4$ 
while it evolves continuously at $\bar{T}=1.5$, suggesting a second order phase transition in the latter. 

The density fields at a given cross-section for various temperatures and for relative 
humidity $h=\exp\left(\left(\mu-\mu^{3D}_{sat}\right)/k_{B}T\right)=0.96$ are shown in 
Figs. 3(a)-3(d) visualizing the 
extent of the diffusive interface, increasing as $\bar{T}\rightarrow\bar{T}_{cc}$. 
Furthermore, the density distributions at a given $h$ show a bimodal response for $\bar{T}\ll\bar{T}_{cc}$ 
as expected for a first order phase transition while its bimodality progressively disappears 
with increasing $\bar{T}$, i.e. temperature as control parameter, 
a hallmark of a second order phase transition (see Figs. 3(e)-3(h)).

\begin{figure}[b!]
\centering
\includegraphics[trim={0.60cm 0.2cm 0.2cm 0.3cm},clip,width=3.4in]{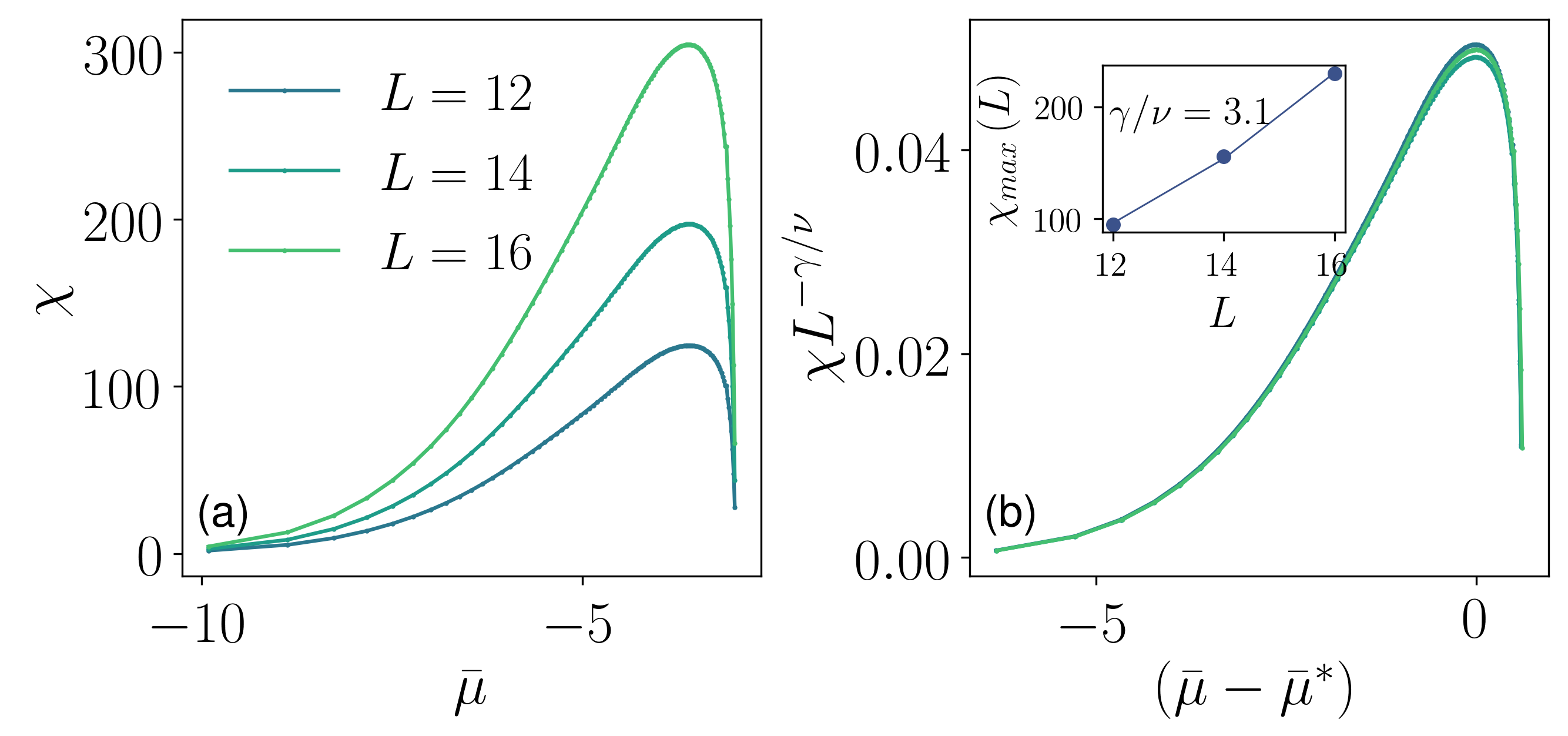}
\caption{(a) Susceptibility for structure $C$ and $\bar{T}=1.5$ for volumes of 
length $L=12$ ($N$=217), $L=14$ ($N$=126) and $L=16$ ($N$=125). 
(b)  Collapse of the susceptibility curves (inset: a power law fit to estimate $\gamma$).}
\label{fig:5}
\end{figure}

Capillary pressure is a manifestation of phase coexistence. 
Capillary curves describe the relationship between liquid saturation $s=\langle\rho\rangle_{\Omega_{p}}$ 
and capillary pressure $p_{c}\left(s\right)=\langle p\rangle_{\Omega_{g}}-\langle p\rangle_{\Omega_{l}}$ where 
$\Omega_{g}$ and $\Omega_{l}$ denote gas and liquid domains, respectively \cite{Charlaix2009}. 
Capillary pressure can be estimated from Eq. \eqref{eq2} via the first moment of pressure in 
the liquid domain, $p_{c}\approx -\langle p\rangle_{\Omega_{l}}$, given that 
$\langle p\rangle_{\Omega_{g}}$ in the gas domain is relatively negligible. 
The liquid domain $\Omega_{l}$ is determined via a threshold for local density $\rho\left(\vec{x}_{i}\right)$ 
with the results for $\rho_{\text{th.}}=0.55$ shown in Figs. 2(e)-2(h). 
The choice of local density threshold does not impact capillary curves significantly (see Supplemental Material \cite{SIPRL2020}). 
These curves exhibit two distinct regimes: a sharp decrease with $s$ associated 
with the buildup of adsorbed film on the pore-solid surfaces, followed by a smooth decrease 
for $\bar{T}\ll\bar{T}_{cc}$. In the vicinity of $\bar{T}\approx\bar{T}_{cc}$, 
the capillary curves suggest pore filling/emptying at zero capillary pressure with 
the first regime absent, signaling the termination of phase coexistence (Figs. 2(e)-2(h)). 
This can also be observed for CP (see Supplemental Material \cite{SIPRL2020}). 
Furthermore, both the isotherms and capillary curves show no particular dependence on the 
degree of spatial disorder although they do display a pronounced 
dependence on $f_{s}$ as the behavior pertaining to structure $D$ consistently differs from structures $A-C$, 
given that all studied structures, $f_{s}\in\{0.43,0.5\}$, can be classified as dilute suspensions. 
However, the higher order cumulants of the capillary pressure fields are sensitive to spatial disorder (see Supplemental Material \cite{SIPRL2020}).

To further explore the nature of capillary phase transition, we carry out finite-size scaling (FSS) analysis, 
and the critical exponents $\nu$ and $\gamma$ governing singularities in correlation length and 
connected susceptibility
are determined for both PS and GM. 
To this end, connected susceptibility $\chi=L^{3}\left(\langle\rho^{2}\rangle-\langle\rho\rangle^{2}\right)$ 
is computed for volumes of characteristic length $L$ chosen to provide a relatively large 
number of realizations ($N>100$) based on the diameter of the particles (pores). For each realization $x$, $\chi_{max}\left(x,L\right)$ is obtained. 
From $\chi_{max}\left(L\right)=\langle\chi_{max}\left(x,L\right)\rangle$ and 
its corresponding chemical potential 
$\bar{\mu}^{*}\left(L\right)=\langle\bar{\mu}^{*}\left(x,L\right)\rangle$, the critical exponents 
are estimated and reported in Tab. \ref{Tab1}. 
\begin{table}[b]
\caption{\label{Tab1}%
Critical exponents estimated from FSS.
}
\begin{ruledtabular}
\begin{tabular}{lcccc}
\textrm{}&
\textrm{$A$}&
\textrm{$B$}&
\textrm{$C$}&
\textrm{$D$}\\
\colrule
$\left(\nu,\gamma\right)_{\text{GM}}$ & $\left(0.68,2.14\right)$ & $\left(0.77,2.43\right)$ & $\left(0.88,2.89\right)$ & $\left(0.81,2.54\right)$\\
$\left(\nu,\gamma\right)_{\text{PS}}$ & $\left(0.76,2.21\right)$ & $\left(0.82,2.42\right)$ & $\left(0.94,2.70\right)$ & $\left(0.47,1.39\right)$
\end{tabular}
\end{ruledtabular}
\end{table}
The quality of the fits are reasonable (see Supplemental Material \cite{SIPRL2020}) but 
obviously the accuracy increases with a larger number of realizations and a larger set of coarse-graining 
lengths that span at least a decade. The obtained values of critical exponents lead to a 
reasonable collapse for the susceptibility curves with an example shown in Fig. 4. 

With regards to the nature of phase transition near $\bar{T}_{cc}$, our estimations 
for $\nu$ suggest a second order phase transition given its discrepancy with the expected 
scaling for a first order transition in a mean-field theory, i.e. $\nu\approx 2/d\left(=3\right)$ \cite{Kierlik2002}. 
This observation combined with the disappearance of bimodality of the 
density distribution, at a given $h$, as temperature increases, and the continuous evolution of density 
near $T_{cc}$ suggest a second order phase transition near $T_{cc}$. 
Furthermore, our results for $\nu$ and $\gamma$ are within the range of those 
reported in the literature for 3D-RFIM for a variety of idealized random fields 
and consistent with universality of confined colloid-polymer mixture \cite{Vink2006,Vink2008} 
with $\nu=1.1\pm0.1$ and $\gamma=2.02\pm0.49$. 
Additionally, the reported critical exponents, $\nu$ and $\gamma$, for 3D-RFIM with an 
underlying Gaussian distribution are $\nu\in[0.96,1.46]$ and $\gamma\in[1.7,2.51]$ \cite{Ahrens2013,Sourlas2014,Machta2006,Sourlas2015,Rieger1995,Sourlas1997,Barkema1996,Fytas2016}, 
for a double Gaussian distribution $\nu\in[1.33,2.68]$ and $\gamma\in[1.98,4.0]$, 
a Poisson distribution $\nu=1.31\pm0.08$ and $\gamma=1.95\pm 0.12$ \cite{Fytas2016}. 
Given the limited options for coarse-graining lengths and the inherent challenges in 
extracting critical exponents \cite{Vink2008}, the agreements between the reported results 
with those in the literature are very promising. Moreover, we observe that 
even for the ordered structure, $A$, with periodic arrangement of particles (pores), the critical exponents 
are in agreement with those reported for 3D-RFIM. This is consistent with de Gennes's 
conjecture as the underlying random field is generated by the distribution of wall separation 
best captured by pair distribution functions shown in Fig. 1(d), highlighting the 
local disorder in fluid-solid interactions.  For the isolated spherical voids in the PS, the associated 
pair distance distribution functions are Gaussians \cite{Koch2003} and hence the agreement 
between the values reported in Tab. \ref{Tab1} and those in the literature with underlying Gaussian random fields.

To conclude, we demonstrated that confinement effects are much less pronounced 
in the studied granular media as opposed to their porous solid counterparts. 
This was shown to be a consequence of the surface-surface correlation length with a 
connected path through the fluid domain as captured via the function $N_{s}^{f}\left(r\right)$. 
In granular aggregates, this correlation length approaches that of the bulk fluid, 
recovering a bulk fluid behavior. 
At the same time, critical exponents estimated from FSS analysis map 
both GM and PS into the 3D-RFIM as previously hypothesized by de Gennes \cite{deGennes1984}. 
This implies that the universality class can be resolved in absence of strong confinement 
with the underlying effective random field being a consequence of local disorder 
in fluid-solid interactions captured by their pair distribution function and the associated 
pair distance distribution function in the pore domain and not necessarily the spatial 
arrangement of the particles (pores). Furthermore, our results suggest a first order phase 
transition for $T\ll T_{cc}$ and a second order phase transition for $T\approx T_{cc}$ 
irrespective of the degree of disorder and the nature of solid matrix, whether discrete or continuous. 
This is based on the estimations for critical exponent $\nu$, evolution of isotherms, 
capillary pressure evolution with temperature and the distribution of density fields. 
Additionally, from the capillary curves, the termination of phase coexistence occurs at $T\approx T_{cc}$. 
This implies that $T_{cc}$ represents a true critical temperature that is insensitive to 
the degree of disorder and the nature of solid matrix. 

In the future, the critical behavior of random porous materials should be examined 
beyond the dilute suspension limit with a stronger degree of heterogeneity, 
e.g. effective random fields with underlying L\'{e}vy stable distributions \cite{Levy1995} accounting for chemical disorder,  i.e. spatially varying fluid-solid interactions, and including correlated structures. The scaling properties of the hull of percolation \cite{Sapoval1985,Rosso1986,Debierre1994} 
can be illuminating in exploring surface-surface correlations in more complex pore domains. 
Lastly, the role of solid deformability on the nature of liquid-gas phase transition remains to be explored. 

\begin{acknowledgments}
The authors would like to thank Prof. Mehran Kardar (Department of Physics at MIT) for reviewing the first draft of this paper and providing very insightful suggestions including carrying out a finite-size scaling analysis. The authors also thank Prof. Emanuela Del Gado (Department of Physics at Georgetown University) for reviewing the last draft of this paper before its final version and providing critical feedback incorporated into this final version. S.M. and T.Z. also thank Prof. Enrico Masoero (School of Engineering at Newcastle University) for fruitful discussions. T.Z. thanks Drinkward fellowship at Caltech's Mechanical \& Civil Engineering Department.  
\end{acknowledgments}

\bibliography{capillary_criticality_arXiv_08102020.bib}
\end{document}